\documentclass[twocolumn,showpacs,aps,prl,superscriptaddress]{revtex4}

\usepackage{graphicx}
\usepackage{dcolumn}
\usepackage{amsmath}
\usepackage{epsfig}
\usepackage{relsize}

\RequirePackage{xspace}

\newcommand {\brhoenu}{$\ensuremath{B^+\rightarrow\rho^0 e^+\nu}$}
\newcommand {\brhochgenu}{$\ensuremath{B^0\rightarrow\rho^- e^+\nu}$}
\newcommand {\bpienu}{$\ensuremath{B^+\rightarrow\pi^0 e^+\nu}$}
\newcommand {\bpichgenu}{$\ensuremath{B^0\rightarrow\pi^- e^+\nu}$}
\newcommand {\bomegaenu}{$\ensuremath{B^+\rightarrow\omega e^+\nu}$}

\newcommand{\BABARPubYear}    {02}
\newcommand{\BABARPubNumber}  {15}

\newcommand{\SLACPubNumber} {9618}

\def\babar{\mbox{\slshape B\kern-0.1em{\smaller A}\kern-0.1em
    B\kern-0.1em{\smaller A\kern-0.2em R}}}
\def\pep2{PEP-II}
\def\epem{\ensuremath{e^+e^-}\xspace}
\mathchardef\Upsilon="7107
\def\Y#1S{\ensuremath{\Upsilon{(#1S)}}\xspace}
\def\FourS{\Y4S}
\def\invfb{\ensuremath{\mbox{\,fb}^{-1}}\xspace}
\newcommand{\mev}{\ensuremath{\mathrm{\,Me\kern -0.1em V}}\xspace}

\def\figurebox#1#2#3{%
    \def\arg{#3}%
    \ifx\arg\empty
    {\hfill\vbox{\hsize#2\hrule\hbox to #2{\vrule\hfill\vbox to #1{\hsize#2\vfill}\vrule}\hrule}\hfill}%
    \else
    {\hfill\epsfbox{#3}\hfill}%
    \fi}

\begin{document}

\preprint{\babar-PUB-\BABARPubYear/\BABARPubNumber} 
\preprint{SLAC-PUB-\SLACPubNumber} 

\begin{flushleft}
\babar-PUB-\BABARPubYear/\BABARPubNumber\\
SLAC-PUB-\SLACPubNumber
\end{flushleft}

\title{
{\large \bf Measurement of the CKM Matrix Element $|V_{ub}|$
with $B \rightarrow \rho e \nu$ Decays}}

%
\author{B.~Aubert}
\author{R.~Barate}
\author{D.~Boutigny}
\author{J.-M.~Gaillard}
\author{A.~Hicheur}
\author{Y.~Karyotakis}
\author{J.~P.~Lees}
\author{P.~Robbe}
\author{V.~Tisserand}
\author{A.~Zghiche}
\affiliation{Laboratoire de Physique des Particules, F-74941 Annecy-le-Vieux, France }
% ...
\collaboration{The \babar\ Collaboration}
\noaffiliation

\date{\today}

\begin{abstract}

  We present a measurement of the branching fraction for the rare
  decays $B \rightarrow \rho e \nu$ and extract a value for the
  magnitude of $V_{ub}$, one of the smallest elements of the
  Cabibbo-Kobayashi-Maskawa quark-mixing matrix. The results are given
  for five different calculations of form factors used to parametrize
  the hadronic current in semileptonic decays. Using a sample of $55$
  million $B\bar{B}$ meson pairs recorded with the \babar\ detector at
  the \pep2 \epem storage ring, we obtain ${\cal
  B}(B^0\rightarrow\,\rho^-\,e^+\,\nu)\,=\,
  (3.29\,\pm\,0.42\,\pm\,{0.47}\pm\,0.60)\,\times\,10^{-4}$ and
  $|V_{ub}|\,=\,(3.64\,\pm\,0.22\,\pm\,0.25\,{}^{+0.39}_{-0.56})\,\times\,10^{-3}$,
  where the uncertainties are statistical, systematic, and
  theoretical, respectively.

\end{abstract}


\maketitle

Exclusive $b \rightarrow u \ell \nu$ decays can be used to determine
$|V_{ub}|$, one of the smallest and least well-determined elements of the
Cabibbo-Kobayashi-Maskawa quark-mixing matrix. 
The modes $B \rightarrow \rho e \nu$ have a comparatively large
branching fraction, and a high fraction of events is found at large
electron momenta.
We determine both the branching fraction ${\cal B}(B \rightarrow \rho
e \nu)$ and $|V_{ub}|$ using form factors, which describe the hadronic
current in the decay, to extrapolate the decay rates to the full range
of lepton energies and to normalize ${\cal B}$ to $|V_{ub}|$. Five
different form-factor calculations are used, as given in
Table~\ref{tab:gamth}.

The data in this analysis were collected with the \babar\
detector~\cite{nim} at the \pep2~\cite{pepii} asymmetric-energy
$\epem$ storage ring. The integrated luminosity of the sample recorded
on the \FourS\ resonance in years $2000$ and $2001$ (``on-resonance'')
is $50.5\invfb$, corresponding to $55.2$ million $B\bar{B}$ meson
pairs. An additional $8.7\invfb$ of data were taken $40\mev$ below the
resonance (``off-resonance''). \babar\ is a detector optimized for the
asymmetric beam configuration at \pep2. Charged-particle momenta are
measured in a tracking system consisting of a 5-layer, double-sided
silicon vertex tracker (SVT) and a 40-layer drift chamber (DCH) filled
with a mixture of helium and isobutane, both operating in a $1.5$-T
superconducting solenoid. The electromagnetic calorimeter (EMC)
consists of 6580 CsI(Tl) crystals arranged in barrel and forward
endcap subdetectors. Particle identification is performed by combining
information from ionization measurements in the SVT and DCH, energy
deposits in the EMC, and the angle and number of Cherenkov photons
measured by the DIRC (detector of internally reflected Cherenkov
light).

We select decays in the modes \brhoenu,\; \brhochgenu,\; \bomegaenu,\;
\bpienu,\; and \bpichgenu\;, with $\rho^0 \rightarrow \pi^+\pi^-$,
$\rho^- \rightarrow \pi^0\pi^-$, and $\omega \rightarrow
\pi^0\pi^+\pi^-$. The inclusion of charge conjugate decays is implied
throughout. The analysis is optimized for $B \rightarrow \rho e \nu$
decays, similar to that in Ref.~\cite{dlange}. Each signal event is
sometimes reconstructed in one of the four other modes; the $\pi$ and
$\omega$ modes are included in order to estimate this crossfeed into
the $\rho$ modes. Throughout this paper, all variables are expressed
in the $\Upsilon(4S)$ center-of-mass frame, except if stated
otherwise. Two electron-energy regions are considered: $2.0\le
E_e<2.3\mbox{ GeV}$ ({\it low-$E_e$}) and $2.3\le E_e<2.7\mbox{ GeV}$
({\it high-$E_e$}). A large background to $b\rightarrow u e\nu$ decays
comes from the more copious $b\rightarrow c e\nu$ decays. This
background is kinematically suppressed in the high-$E_e$ region and
dominates in the low-$E_e$ region. The low-$E_e$ region provides the
background normalization in the high-$E_e$ region. The largest
background in the high-$E_e$ region is continuum $e^+e^-\rightarrow
q\bar{q}$ events. The off-resonance data are used to estimate its
size.

Hadronic events are selected based on track and photon multiplicity
and event topology. We use tracks originating from the interaction
point with at least 12 hits in the DCH and a transverse momentum
greater than $0.1\mbox{ GeV}/c$.  Signals in the EMC with
$E_{lab}>30\mbox{ MeV}$ that are not associated with any track are
considered as photons if the lateral moment of the shower energy
distribution~\cite{LAT} is smaller than $0.8$. We select events with
at least five tracks, or with at least four tracks and at least five
photons. We require the ratio $H_2/H_0$ of Fox-Wolfram
moments~\cite{foxw} to be less than~0.4. This requirement keeps 85\%
of the $\rho e \nu$ signal; it rejects 55\% of the non-$B\bar{B}$
events.

\begin{table}[t]
\caption{Form-factor calculations used in the determination of ${\cal
B}(B\rightarrow\rho e \nu)$ and $|V_{ub}|$, and predicted
normalizations $\tilde{\Gamma}_{{\rm th}}$ (as defined later in
Eq.~\ref{eq:vub}).}
\begin{center}
\begin{tabular}{lccc} \hline \hline
Form factors     & $\tilde{\Gamma}_{{\rm th}}$ (ps$^{-1}$) & Error (\%) & Reference \\ \hline
ISGW2            & 14.2 & $\pm 50$         & \cite{isgw2}\\
Beyer/Melikhov   & 16.0 & $\pm 15$         & \cite{beyer98}\\ 
UKQCD            & 16.5 & $+21,-14$        & \cite{ukqcd}\\
LCSR             & 16.9 & $\pm 32$         & \cite{lcsr}\\
Ligeti/Wise      & 19.4 & $\pm 29$         & \cite{ligeti}\\ \hline \hline
\end{tabular}
\end{center}
\label{tab:gamth}
\end{table}
 
Electrons are identified with a likelihood estimator using information
from the DCH, EMC, and DIRC subdetectors~\cite{inclbabar}. The
selection efficiency is around $90\%$, with a pion misidentification
rate of less than $0.1\%$. We reject electrons from $J/\psi$ decays
and from photon conversions.

Charged pion candidates are tracks not identified as kaons
with high confidence based on DIRC and $dE/dx$ measurements. A
$\pi^0$ is reconstructed from photon pairs with an invariant mass $120
< M_{\gamma\gamma} < 145\mbox{ MeV}/c^2$.

To reconstruct $\rho^0$ mesons, we combine two oppositely-charged
pions, and for $\rho^\pm$ a pion track and a $\pi^0$. To suppress
combinatorial background we require that the pion with the higher
momentum satisfies $p_{\pi}>400\mbox{ MeV}/c$ and the other pion
$p_\pi>200\mbox{ MeV}/c$. For the $\omega$, we combine two
oppositely-charged pions with a $\pi^0$. To suppress combinatorial
background we require $p_\pi>100\mbox{ MeV}/c$ for each pion. In the
mode $B \rightarrow \pi e \nu$ we require $p_\pi>200 \mbox{ MeV}/c$.

The missing momentum in the event is given by
\begin{equation}
\vec{p}_{\rm miss} = - \sum_{\rm tracks}{\vec{p}_i} - \sum_{\rm photons}{\vec{p}_i}\;,
\end{equation}
where the sums are over all accepted tracks and photons. We require
$\vert \cos \theta_{{\rm miss}} \vert<0.9$, where $\theta_{\rm miss}$
is the angle between $\vec{p}_{\rm miss}$ and the beam axis. This
rejects events with missing high-momentum particles close to the beam
axis. We also compare the direction of $\vec{p}_{\rm miss}$ with that
of the neutrino inferred from $\vec{p}_{\nu} =
\vec{p}_{B}-\vec{p}_{Y}$, where $Y$ is the $\rho + e$, $\omega + e$,
or $\pi + e$ system. The latter is known to within an azimuthal
ambiguity about the $B$ direction since only the magnitude of
$\vec{p}_{B}$ is known. We use the smallest possible angle
$\Delta\theta_{\rm min}$ between the two directions and require $\cos
\Delta\theta_{\rm min}>0.8$. Using the constraints $E_{B} = E_{\rm
beam}$ and $p_{\nu}^2=(p_B - p_Y)^2 = 0$, the angle between the $B$
meson and the $Y$ system is
\begin{equation}
\cos \theta_{BY}=\frac{2 E_B E_{Y} - (M_B^2  + M_{Y}^2) c^4} 
                      {2 |\vec{p}_B| |\vec{p}_{Y}| c^2}\;.
\label{eq:cosby}
\end{equation}
Signal events fulfill $|\cos \theta_{BY}| \leq 1$; allowing for detector
resolution we require $|\cos \theta_{BY}|<1.1$.
After all other selection criteria, this requirement rejects more than
60\% of the $b \rightarrow c e \nu$ and approximately 68\% of the
remaining continuum backgrounds, it retains 98\% of the signal.

To further reduce the continuum background, we use a neural net with
14 event-shape variables: the sum of track and photon energies in nine
cones centered on the lepton-momentum; the angle $\theta_{{\rm
thrust}}$ between the thrust axis of the $Y$ system and the thrust
axis of the rest of the event (the thrust axis is defined to be the
direction that maximizes the sum of the longitudinal momenta of all
particles); the angle $\theta_{{\rm thrust},Y}$ between the thrust of
the $Y$ system and the beam axis; the angle $\theta_{{\rm lept,rest}}$
between the direction of the lepton and the direction of the total
momentum of all tracks except the $Y$ system; the momentum of the
track with the smallest opening angle with respect to the electron;
$\sum_i \vec{p}_i \cdot \vec{n}_{e}/\sum_i \vert \vec{p}_i \vert$,
where $\vec{n}_{e}$ is the direction of the electron and $\vec{p}_i$
are the momenta of all tracks except the electron. After all other
selection criteria, the neural net condition removes more than 90\% of
the continuum events in the high-$E_e$ region, while retaining
approximately 60\% of the signal events in each signal mode.

After all selections, there remain on average $3.4$ candidates per
event. We choose the one with a total momentum $|\vec{p}_Y +
\vec{p}_{\rm miss}|$ closest to the $B$-meson momentum
$|\vec{p}_B|$. The probability of making the right choice for the
signal modes is approximately 85\%.

The total efficiency in the high-$E_e$ region is $12.0\%$ ($9.5\%$)
for the mode $B^+ \rightarrow \rho^0 e^+ \nu$ ($B^0 \rightarrow \rho^-
e^+ \nu$) in the ISGW2 model; it is $4.2\%$ ($3.3\%$), when
relating the accepted events in the high-$E_e$ region to 
events with all electron energies.

We perform a binned maximum-likelihood fit to the two-dimensional
distribution ($M_{\pi \pi (\pi)}$, $\Delta E$), where $M_{\pi \pi
(\pi)}$ is the invariant mass of the $\rho$ ($\omega$) meson
and $\Delta E$ is the difference between the reconstructed and the
expected $B$-meson energy, $\Delta E \equiv E_{\rm hadron} + E_{e}
+ \vert {\vec{p}}_{{\rm miss}} \vert c - E_{\rm beam}$. The fit is
performed simultaneously for the five signal modes 
in the two $E_e$ ranges. For the $B \to \rho e
\nu$ modes, the data are divided into $10\times10$ bins over the
($M_{\pi\pi}$, $\Delta E$) region $0.25 \leq M_{\pi\pi} \leq
2.00\mbox{ GeV}/c^2$ and $|\Delta E| \leq 2\mbox{ GeV}$. For the
$\omega$ channel, we use five bins in the range $702 \leq M_{\pi\pi\pi}
\leq 862 \mbox{ MeV}/c^2$ and ten bins in $|\Delta E|\leq 2\mbox{
GeV}$. For the modes $B \rightarrow \pi e \nu$, only $\Delta E$ 
is used as a fit variable, also with ten bins.

In the fit, the likelihood is calculated as product of probability
distributions for each of the five signal modes, for other $b \to u e
\nu$ decays, for $b \to c e \nu$ decays, for continuum events, and for
a small contribution due to misidentified electrons. Shapes and
normalizations of the continuum background and misidentified electrons
are extracted from the data. For all other contributions, Monte Carlo
(MC) simulation provides the shapes of the distributions. The decays
$B\to D^{(*)}e\nu$ are simulated using a model based on heavy quark
effective theory~\cite{HQET}. The modes $B\to D^{(*)}\pi e\nu$ are
simulated according to the Goity-Roberts model~\cite{goity}. The
resonances $b \to u e \nu$ heavier than $\rho$ and $\omega$ are
implemented according to the ISGW2 model~\cite{isgw2}. Non-resonant $b
\to u e \nu$ modes are described by the model of Fazio and
Neubert~\cite{Xu}.

The fit has nine free parameters: ${\cal B}(B^0 \rightarrow
\rho^- e^+\nu)$, ${\cal B}(B^0\rightarrow\pi^-e^+\nu)$, the
normalization of the $b \rightarrow u e \nu$ background in the two
electron-energy ranges (two parameters), and the normalization of the
$b \to c e \nu$ background (five parameters, one for each mode). 
The rates of the $\rho^0$, $\omega$, and $\pi^0$ channel are
constrained by the isospin and quark model relations
$\Gamma(B^0\rightarrow\rho^-e^+\nu)=2\Gamma(B^+\rightarrow\rho^0 e^+\nu)$,
$\Gamma(B^+\rightarrow\rho^0e^+\nu)=\Gamma(B^+\rightarrow\omega e^+\nu)$, and
$\Gamma(B^0\rightarrow\pi^-e^+\nu)=2\Gamma(B^+\rightarrow\pi^0 e^+\nu)$.
The maximum-likelihood fit takes into account the statistical
uncertainties in the on- and off-resonance data and in the probability
distributions extracted from MC simulations~\cite{barlow}.

\begin{figure}[tb]
  \unitlength1cm
  \begin{minipage}[t]{4.0cm}
    \begin{center}
    \epsfig{file=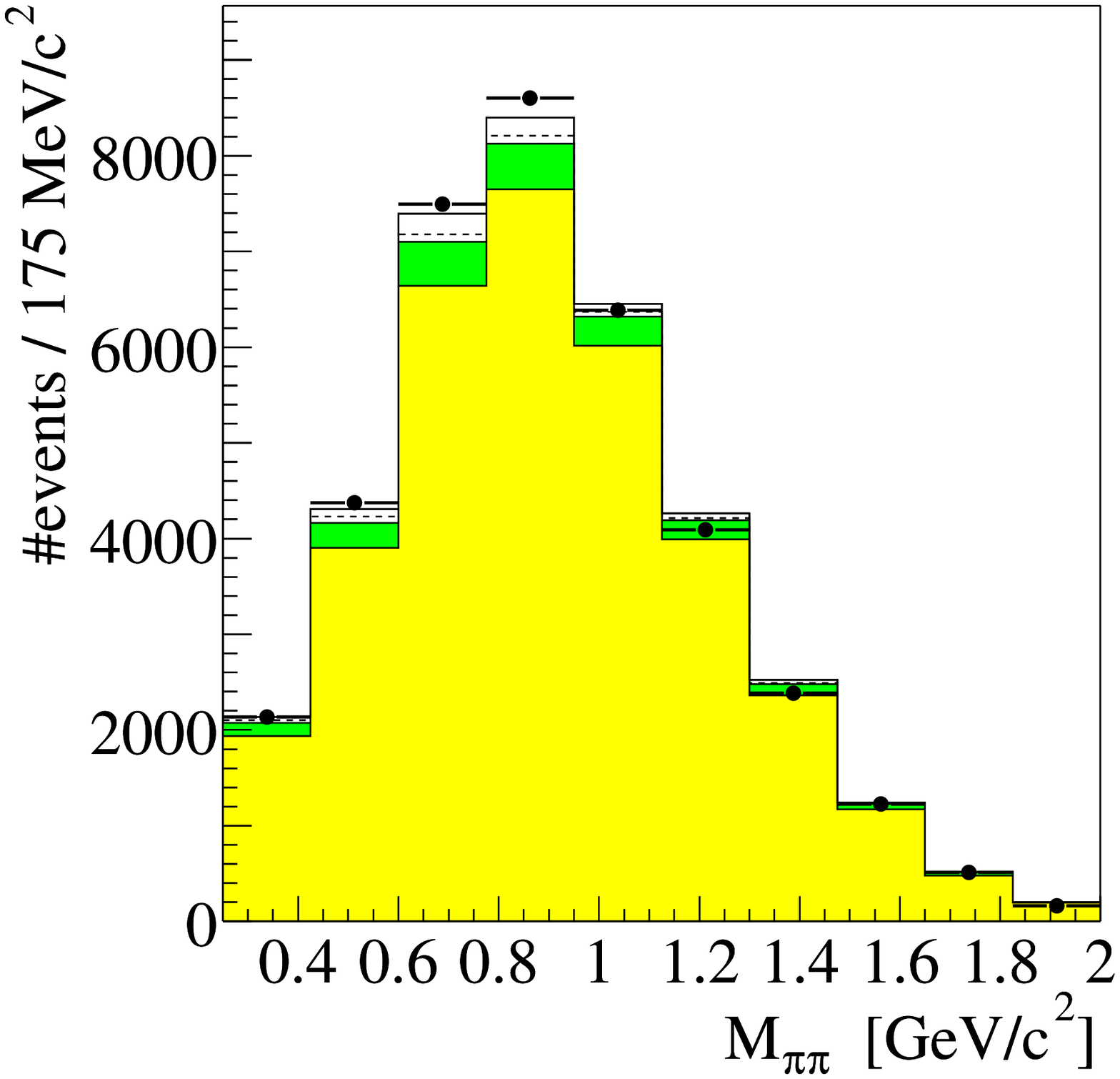,width=4cm,bbllx=20,bblly=5,bburx=550,bbury=520,clip=}
    \put(-1.5,3.2){low-$E_e$}
    \end{center}
  \end{minipage}
  \begin{minipage}[t]{4.5cm}
    \begin{center}
    \epsfig{file=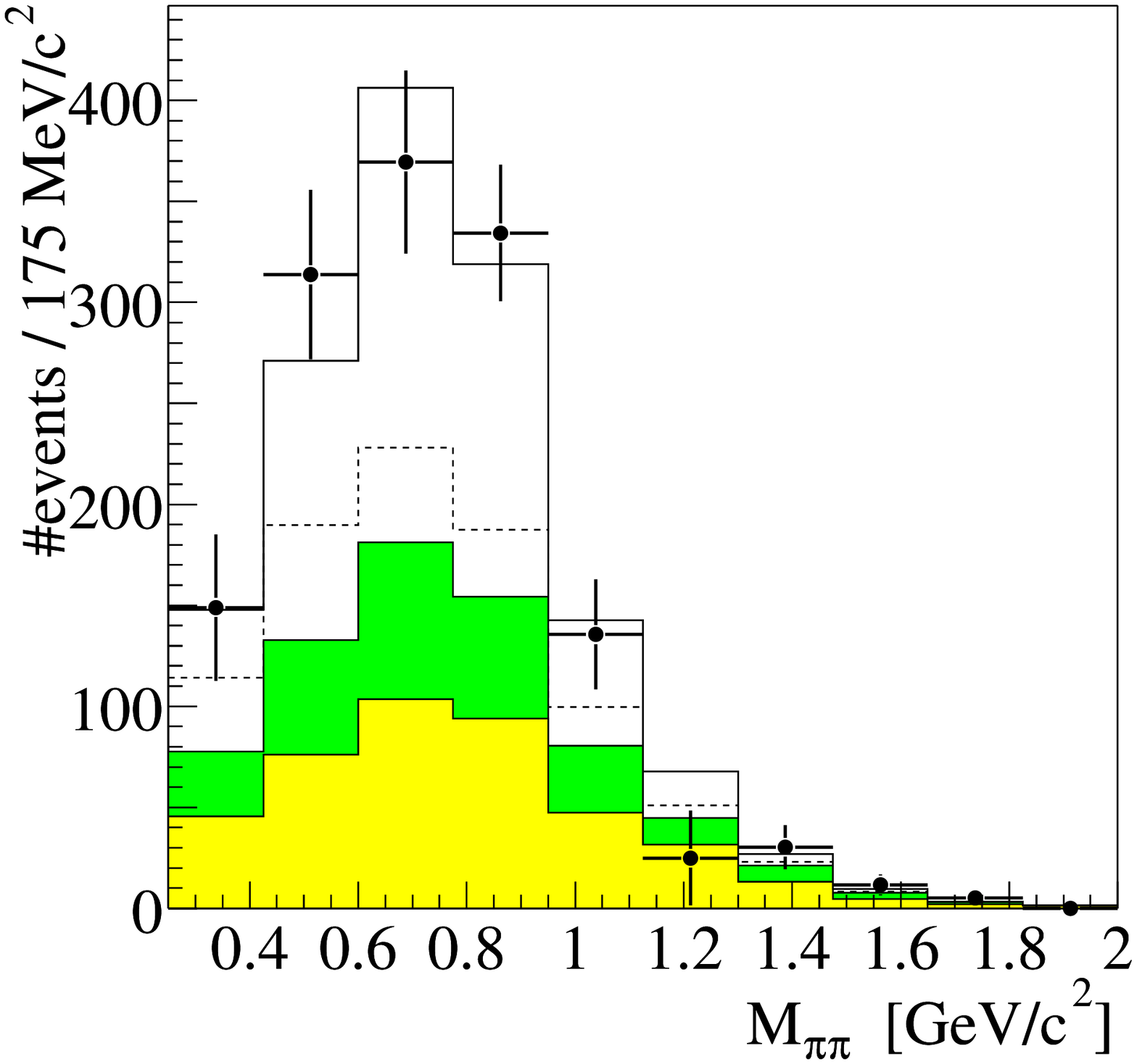,width=4.1cm,bbllx=6,bblly=5,bburx=552,bbury=520,clip=}
    \put(-1.5,3.2){high-$E_e$}
    \end{center}
  \end{minipage}
  \begin{minipage}[t]{4.0cm}
    \begin{center}
    \epsfig{file=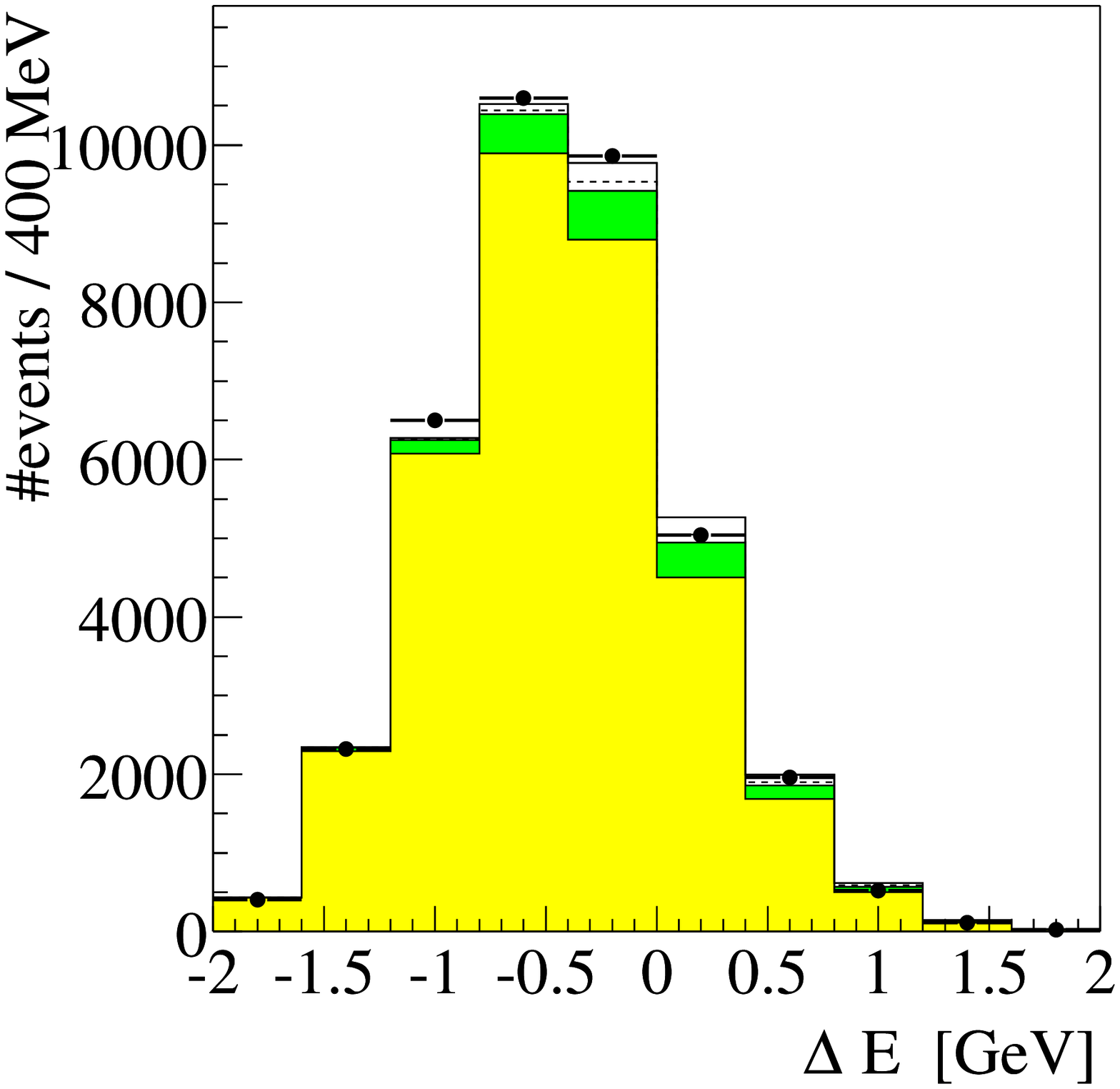,width=4cm,bbllx=20,bblly=5,bburx=550,bbury=520,clip=}
    \put(-1.5,3.2){low-$E_e$}
    \end{center}
  \end{minipage}
  \begin{minipage}[t]{4.5cm}
    \begin{center}
    \epsfig{file=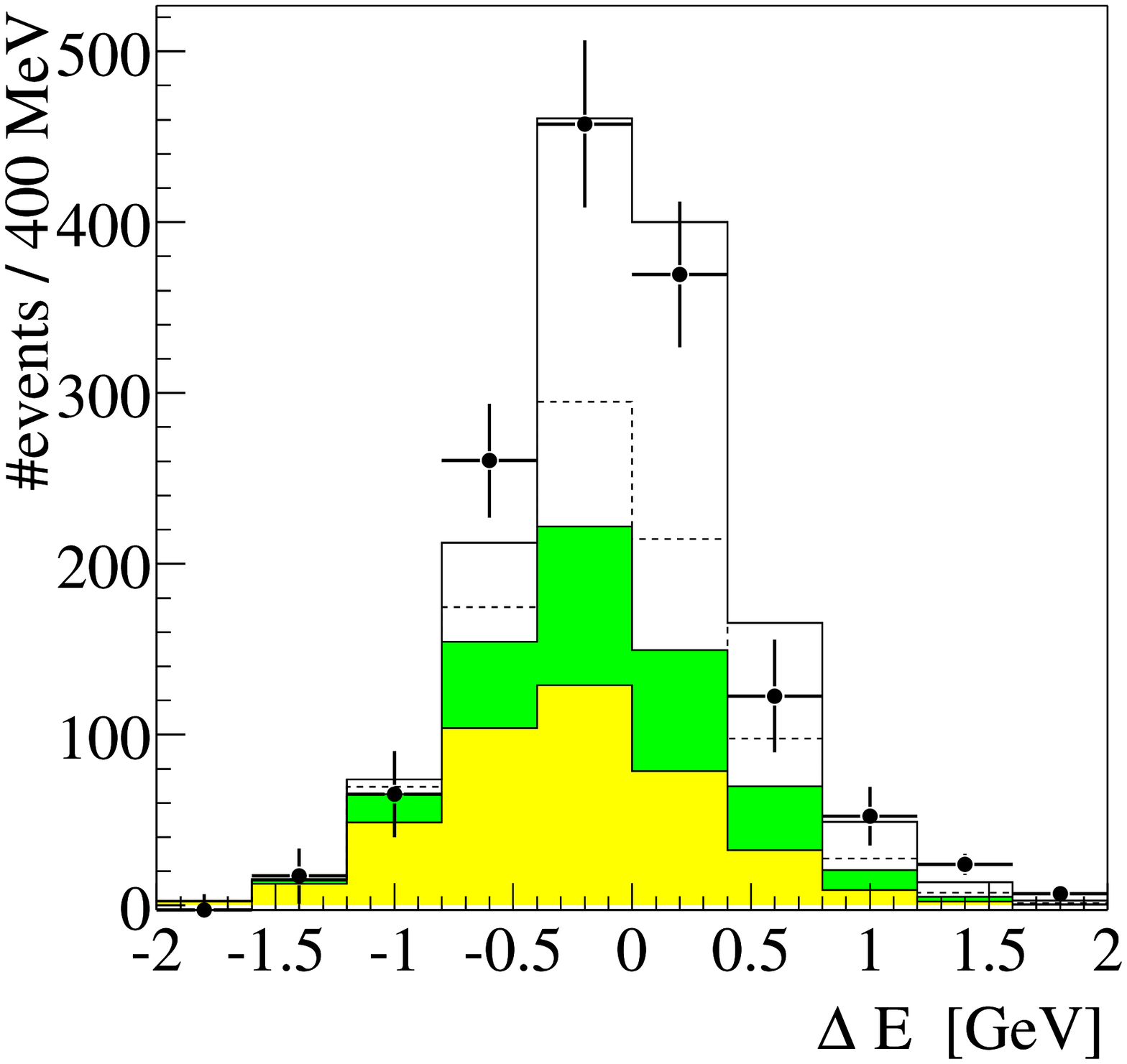,width=4.1cm,bbllx=6,bblly=5,bburx=552,bbury=520,clip=}
    \put(-1.5,3.2){high-$E_e$}
    \end{center}
  \end{minipage}
\caption{Continuum-subtracted data distributions (points with error
bars) and fit projections (histograms) for $M_{\pi\pi}$ (top plots)
and $\Delta E$ (bottom plots) for the \brhochgenu\; channel in the
low-$E_e$ (left plots) and high-$E_e$ regions (right plots). The fit
results are shown for the ISGW2 model. The histograms
correspond to the true and crossfeed components of the signal (open
histogram, above and below the dashed line, respectively), the
background from other $b\to u e \nu$ decays (dark shaded region), and
$b\to c e \nu$ and other backgrounds (light shaded region).}
\label{fig:resrhochg} 
\end{figure}

\begin{figure}[tb]
\begin{center}
\unitlength1cm
\epsfig{file=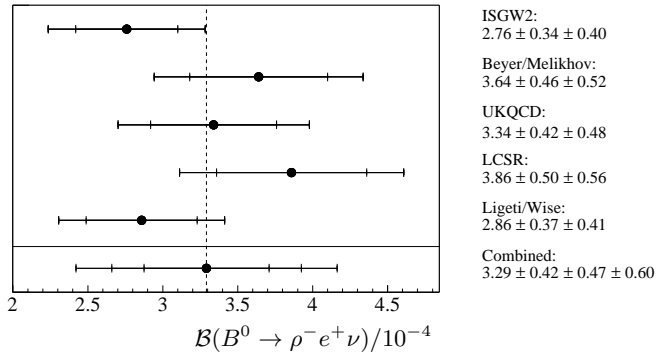,width=9cm}
\put(-6.4,0){${\cal B}(B^0 \rightarrow \rho^- e^+ \nu) / 10^{-4}$}
\end{center}
\caption{The \brhochgenu\; branching fraction results using five
different form-factors calculations. The uncertainties shown are
statistical, systematic, and (for the combined result)
theoretical, successively added in quadrature. The combined
result is the average of the five form-factor results.}
\label{fig:brrho}  
\end{figure}

Projections of the data and fit results for \brhochgenu\; are shown in
Fig.~\ref{fig:resrhochg} for the ISGW2 model. A continuum-background
contribution of $917\pm73$ events in high-$E_e$ and $1928\pm106$ in
low-$E_e$ has been subtracted. Good agreement between data and the fit
result is seen in each of these figures. The fits for the other
form-factor calculations show the same level of agreement.  The fit
quality has been checked with a $\chi^2$ test, where bins in sparsely
populated regions have been combined before the $\chi^2$
calculation. We obtain $\chi^2=91$ for $93$ degrees of freedom for
ISGW2, and similarly good fit quality for the other form-factor
calculations. The signal yields extracted from the maximum-likelihood
fit in the high-$E_e$ region are $321 \pm 40$ \brhoenu\; events and
$505 \pm 63$ \brhochgenu\; events. The resulting branching fractions
${\cal B}(B^0 \rightarrow \rho^-e^+\nu)$ are shown in Fig.~\ref{fig:brrho}. The
five fit parameters describing the $b \rightarrow c e \nu$ backgrounds
agree with the known branching fractions~\cite{pdg2002} for
$B\rightarrow D e\nu$, $B\rightarrow D^* e\nu$, and $B\rightarrow
D^{(*)}(\pi)e\nu$ within $\pm9\%$ on average. The two parameters
describing the size of the background from other $b\rightarrow u e\nu$
decays agree within $1.5\sigma$ with the predictions of the MC
simulation. The fit result for the $\pi$ modes is ${\cal B}(B^0
\rightarrow\pi^-e^+\nu)= (1.86\pm0.56_{stat.})\times10^{-4}$ for the
ISGW2 model.

A summary of all considered systematic uncertainties on ${\cal
B}(B\rightarrow \rho e\nu)$ is given in
Table~\ref{systematicErrors}. The total systematic uncertainty is the
quadratic sum of all individual ones. Note that the statistical
uncertainties in Fig.~\ref{fig:brrho} already include the statistical
uncertainty in the MC predictions. The largest single contribution to
the systematic error arises from the uncertainty in the shape of the
$b \rightarrow u e \nu$ background from events other than the signal
modes. The fraction of $b \rightarrow u e \nu$ background events that
are non-resonant is varied from $0$ to $2/3$ to estimate this
uncertainty. The composition of the resonant component of other $b
\rightarrow u e \nu$ decays has been varied by changing the branching
fractions for individual resonances by $\pm 50\%$, while keeping the
total rate constant. Variations in the $b \rightarrow c e \nu$
composition contribute much less to the total systematic error than
variations in the $b \rightarrow u e \nu$ component. Possible
violations of the isospin and quark model constraints are discussed in
Refs.~\cite{iso} and~\cite{dlange2}. Their contribution to the
systematic error is determined by allowing a $\pm 3\%$
violation. Several fits were performed: fitting without the $\omega$
mode, without the $\pi$ mode (fixing ${\cal B}(B \rightarrow \pi e
\nu)$~\cite{pdg2002}), without the low-$E_e$ region, and with
different binning. We assign a systematic uncertainty for the fit
method as half the largest resulting changes of the fit result.  We
have also varied the most important selection requirements and find
that the changes in ${\cal B}(B \rightarrow \rho e \nu)$ are
consistent with statistical variations as determined by a MC
simulation.

\begin{table}[tb]
\caption{Summary of all contributions to the systematic uncertainty on the branching fraction ${\cal B}(B \rightarrow \rho e \nu)$.}
\begin{center}
\begin{tabular}{l|c} \hline \hline
Contribution                               & $\delta{\cal B}_{\rho}/{\cal B}_{\rho}$ (\%) \\ 
\hline
Tracking efficiency                        & $\pm 5$\\
Tracking resolution                        & $\pm 1$\\
$\pi^0$ efficiency                         & $\pm 5$\\
$\pi^0$ energy scale                       & $\pm 3$\\
\hline
$b\rightarrow c e \nu$ background composition    & $+1.4,-1.7$ \\
Resonant $b\rightarrow u e \nu$ background composition & $+6,-4$ \\
Non-resonant $b\rightarrow u e \nu$ background   & $\pm 9$\\
\hline
$B$ lifetime                               & $\pm 1$ \\
Number of $B\bar{B}$ pairs                 & $\pm 1.6$ \\
Misidentified electrons                    & $<\pm 1$ \\
Electron efficiency                        & $\pm 2$ \\
${\cal B}(\Upsilon(4S)\rightarrow B^+B^-)/{\cal B}(\Upsilon(4S)\rightarrow B^0\bar{B}^0)$ & $<\pm 1$ \\
Isospin and quark model symmetries         & $<\pm 1$ \\
Fit method                                 & $+4,-6$ \\ \hline
Total systematic uncertainty               & $\pm 14.4$ \\
\hline \hline
\end{tabular}
\end{center}
\label{systematicErrors}
\end{table}

\begin{figure}[tb]
\begin{center}
\epsfig{file=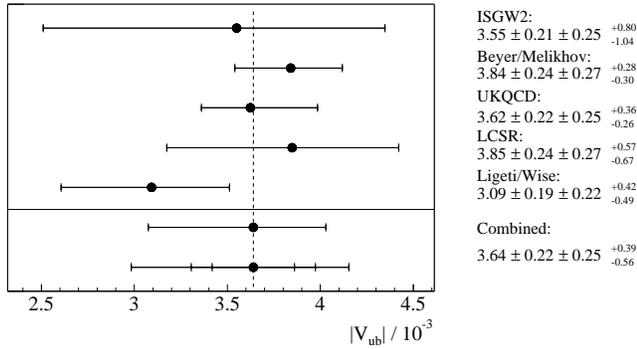,width=9cm}
\end{center}
\caption[$|V_{ub}|$ for different form-factor calculations]{
  $|V_{ub}|$ determined using five different form-factor
  calculations. Only theoretical error bars are shown.
  The combined result is also shown at the bottom with statistical,
  systematic, and theoretical uncertainties successively added in
  quadrature. The combined result is the weighted average of
  the five form-factor results, where we have used only the
  theoretical uncertainties to calculate the weights.}
\label{fig:vub}   
\end{figure}

A value of $|V_{ub}|$ is determined by the relation 
\begin{equation}
\vert V_{ub} \vert = \sqrt{ {\cal B}(B^0\rightarrow
\rho^- e^+ \nu) / {(\tilde{\Gamma}_{{\rm th}} \tau_{B^0})}}\;,
\label{eq:vub}
\end{equation}
where $\tilde{\Gamma}_{{\rm th}}$ is the predicted form-factor
normalization as given in Table~\ref{tab:gamth}. The branching
fractions are used separately for each form-factor calculation, as
shown in Fig.~\ref{fig:brrho}. We use $\tau_{B^0} = 1.542 \pm
0.016$~ps~\cite{pdg2002} for the $B^0$ lifetime. The results for
$|V_{ub}|$ are shown in Fig.~\ref{fig:vub}. The combined result has
been obtained as weighted average of the five form-factor results,
where the weight is obtained from the theoretical uncertainty of each.
The theoretical uncertainty on the combined result is estimated to be
half of the full spread of all theoretical uncertainties.

In conclusion, we have measured the branching fraction
${\cal B}(B^0\rightarrow~\rho^-~e^+~\nu) =
(3.29~\pm~0.42~\pm~{0.47}\pm~0.60)~\times 10^{-4}$
using isospin constraints and extrapolating to all electron energies
according to five different form-factor calculations. The errors given
are statistical, systematic, and theoretical, in the order shown. The
value of $|V_{ub}|$ determined by the same form-factor calculations is
$|V_{ub}| = (3.64~\pm~0.22~\pm~0.25~{}^{+0.39}_{-0.56})~\times~10^{-3}$.
Our results are slightly higher ($22\%$ for $\mathcal{B}$ and $13\%$
for $|V_{ub}|$) than a previous $B\rightarrow\rho e\nu$
result from CLEO~\cite{dlange}, but agree within statistical errors.

We are grateful for the excellent luminosity and machine conditions
provided by our \pep2\ colleagues, 
and for the substantial dedicated effort from
the computing organizations that support \babar.
The collaborating institutions wish to thank 
SLAC for its support and kind hospitality. 
This work is supported by
DOE
and NSF (USA),
NSERC (Canada),
IHEP (China),
CEA and
CNRS-IN2P3
(France),
BMBF and DFG
(Germany),
INFN (Italy),
NFR (Norway),
MIST (Russia), and
PPARC (United Kingdom). 
Individuals have received support from the 
A.~P.~Sloan Foundation, 
Research Corporation,
and Alexander von Humboldt Foundation.

\end{document}